\def\spose#1{\hbox to 0pt{#1\hss}}
\def\approxlt{\mathrel{\spose{\lower 3pt\hbox{$\sim$}}
        \raise 2.0pt\hbox{$$<$$}}}
\def\approxgt{\mathrel{\spose{\lower 3pt\hbox{$\sim$}}
        \raise 2.0pt\hbox{$>$}}}

\def\multleft#1{\hbox to size{\vbox {\halign {\lft{##}\cr #1}}\hfill}\par}
\def\multright#1{\hbox to size{\vbox {\halign {\rt{##}\cr #1}}\hfill}\par}

\def\today{\ifcase\month\or January\or February\or March\or April\or May\or
      June\or July\or August\or September\or October\or November\or December\fi
      \space\number\day, \number\year}
\def\s{\hbox{\phantom{5}}}

\def\boxit#1{\vbox{\hrule\hbox{\vrule\kern3pt\vbox{\kern3pt
          #1 \kern3pt}\kern3pt\vrule}\hrule}}

\def\cm{{\rm\thinspace cm}}

\def\erg{{\rm\thinspace erg}}

\def\keV{{\rm\thinspace keV}}

\def\ph{{\rm\thinspace ph}}
\def\s{{\rm\thinspace s}}

\def\ergpcmsqps{\hbox{$\erg\cm^{-2}\s^{-1}\,$}}

\def\phpcmsqps{\hbox{$\ph\cm^{-2}\s^{-1}\,$}}

\documentstyle[psfig,times]{mn} 
\begin{document}
\hsize=6truein

\def\simless{\mathbin{\lower 3pt\hbox
   {$\rlap{\raise 5pt\hbox{$\char'074$}}\mathchar"7218$}}}   
\def\simgreat{\mathbin{\lower 3pt\hbox
   {$\rlap{\raise 5pt\hbox{$\char'076$}}\mathchar"7218$}}}   
\def\anisotropy{\frac{\Delta T}{T}}

\title[{\it RXTE} observations of MCG$-$6-30-15]{An RXTE Observation of the Seyfert 1 Galaxy MCG$-$6-30-15 : X-ray
  Reflection and the Iron Abundance }

\author[J.~C.~Lee et al.]
{\parbox[]{6.in} { J.C.~Lee,$^1$ A.C.~Fabian$^1$, C.S.~Reynolds,$^2$ K.~Iwasawa,$^1$ and W.N.~Brandt$^3$ \\
\footnotesize \it $^1$ Institute of Astronomy; Madingley Road; Cambridge CB3 0HA \\
\it $^2$ JILA; Campus Box 440; University of Colorado; Boulder, 80309-0440 USA \\ 
\it $^3$ Department of Astronomy and Astrophysics; The Pennsylvania State University; 525 Davey Lab; University Park, PA 16802 USA
}}                                            
\maketitle


\begin{abstract}  
  We report on a 50\,ks observation of the bright Seyfert 1 galaxy
  MCG$-$6-30-15 with the {\it Rossi X-ray Timing Explorer}.  The data
  clearly show the broad fluorescent iron line (equivalent width
  $\sim$ 250 eV), and the Compton reflection continuum at higher
  energies.  A comparison of the iron line and the reflection
  continuum has enabled us to constrain reflective fraction and the
  elemental abundances in the accretion disk.  Temporal studies
  provide evidence that spectral variability is due to changes in both
  the amount of reflection seen and the properties of the  primary
  X-ray source itself.

\end{abstract}


\section{INTRODUCTION} 

The current paradigm for active galactic nuclei (AGN) is a central engine
consisting of an accretion disk surrounding a supermassive black hole
(e.g., see review by Rees 1984). The main source of power is the release of
gravitational potential energy as matter falls towards the central black
hole.  Much of this energy is released in the form of X-rays, some fraction
of which are reprocessed by matter in the AGN (Guilbert \& Rees 1988; Lightman
\& White 1988).

Careful study of the X-ray reprocessing mechanisms can give much
information about the immediate environment of the accreting black hole.
These effects of reprocessing can often be observed in the form of emission
and absorption features in the X-ray spectra of AGNs.  In Seyfert 1
nuclei, approximately half of the X-rays are `reflected' off the inner
regions of the accretion disk. 
Since it is superposed on the direct (power-law) primary X-ray
emission, the principle observables of this reflection are a fluorescent
iron K$\alpha$ line, and a Compton backscattered continuum which hardens
the observed spectrum above $\sim 10\keV$ (see eg. George \& Fabian 1991).
The iron line together with
the reflection component are important diagnostics for the geometry and
physics of the X-ray continuum source. The strength of the emission line
relative to the reflection continuum depends largely on the abundance of iron
relative to hydrogen in the disk, as well as the normalization of the
reflection spectrum relative to the direct spectrum.  
(There is also a dependence on the relative oxygen abundance, Reynolds et al.
1995.) The relative
normalization of the reflection spectrum probably depends primarily on the
geometry (i.e., the solid angle subtended by the reflecting parts of the
disk as seen by the X-ray source).  However, it can also be affected by
strong light bending effects (e.g., Martocchia \& Matt 1996) or
special-relativistic beaming effects (e.g. Reynolds \& Fabian 1997).
Disentangling the abundance from the absolute normalization of the
reflection component is an important first step in constraining these
effects and hence the construction of physical models for AGN central
regions.

MCG$-$6-30-15 is a Seyfert 1 galaxy that is both bright and nearby
(z=0.008).  Since its identification, MCG$-$6-30-15 has been
intensively studied by every major X-ray observatory.  An extended
{\it EXOSAT} observation provided the first evidence for fluorescent
iron line emission (Nandra et al. 1989) which was attributed to X-ray
reflection.  Confirmation of these iron features by {\it Ginga} as
well as the discovery of the associated Compton reflected continuum
supported the reflection picture (Nandra, Pounds \& Stewart 1990;
Pounds et al. 1990; Matsuoka et al. 1990).   ASCA data showed the iron
line to be broad, skewed, and variable (eg. Tanaka et al 1995;
Iwasawa et al. 1997). 

In this paper, we present the first data from the {\it Rossi X-ray
Timing Explorer} ({\it RXTE}) for MCG$-$6-30-15.  Our observation shows
clear evidence for a redshifted broad iron line at $\sim$ 6.1 keV and
the reflection continuum above 10 keV.  Due to the larger effective
area and longer  exposure of the {\it RXTE} observation as compared
with {\it Ginga}, we can study the reflection continuum in detail for
the first time.   We present preliminary constraints on the abundances
of iron and reflective fraction, and investigate the relationship
between spectral changes and the reflection component during the
different  phases of our data.  Section 2 will detail the data
analysis procedure followed by spectral fitting results in Section 3.
We present a study of temporal variations on spectral components with
particular emphasis on the reflection component in Section 4.  This
will follow with a discussion of results and future work in Section 5. 

\section{Observations}

MCG$-$6-30-15 was observed by the {\it RXTE} for 50 ks over the period from
1996 September 15 to 1996 September 25 by both the Proportional Counter
Array (PCA) and High-Energy X-ray Timing Experiment (HEXTE) instruments.
We concentrate on results from the PCA in this paper.

\subsection{Proportional Counter Array}
The {\it RXTE} PCA consists of 5 Xenon Proportional Counter Units (PCUs)
sensitive to X-ray energies between 2-60 keV with $\sim 18 $ per cent energy
resolution at 6 keV.  The total collecting area is 6500 $\rm cm^2$ 
($\sim$ 3900 $\rm cm^2$ for 3 PCUs) with a $1^\circ$ FWHM field of view.  
The HEXTE instrument is coaligned with the PCA and covers an energy 
range between 20-200 keV; these results will be be discussed in a later 
paper.

\subsection{Data Analysis}

We extract PCA light curves and spectra from only the top Xenon layer using
the standard Ftools 4.0 software developed specifically for {\it RXTE}. This was
done to improve the signal-to-noise since the top layer detects $\sim 90 $
per cent of the cosmic photons and $\sim 50 $ per cent of the internal 
instrumental background.  At the expense of slightly blurring the spectral 
resolution, we also combine data from three of the five PCUs in order to 
improve signal-to-noise.   Data from the remaining PCUs (PCU 3 and 4) were 
excluded due to the fact that these instruments periodically suffer discharge
and are hence sometimes turned off.

Good time intervals were selected to exclude any earth or South 
Atlantic Anomaly (SAA) passage occultations, and to ensure stable pointing.
We also require that data from only PCUs 0, 1, and 2 were used.

We generate our background data using {\sc pcabackest v1.5} in order
to estimate the internal background caused by interactions between the
radiation / particles and the detector / spacecraft at the time of
observation.  This is done by matching the conditions of observations
with those in various model files.  The model files that we chose were
constructed using the VLE rate (one of the rates in PCA Standard 2
science array data that is defined to be the rate of events which
saturate the analog electronics) as the tracer of the particle
background. Other models such as the Q6 (short for Standard 2
telemetry rate) background model exist and is based on the rate of events
which trigger exactly six of the eight signal channels.
Both models are believed to be  a measure of the instantaneous
particle flux (Keith Jahoda, 1997 private communication). While the
choice between these two types of models is based largely on value
judgements (neither is perfect at present), we find that the VLE
background models provide a better estimate of the background in a
test using Earth occultation data from our observations. 

There was no need to apply a deadtime correction for our PCA data
since deadtime exceeds 1 per cent only when the count rate per PCU is
greater than 1000 $\rm cts$ $\rm s^{-1}$.  The 3 PCU count rate for
MCG$-$6-30-15  is $\sim 5-30$ $\rm cts$ $\rm s^{-1}$ for the the
background subtracted source (Fig.~1),  and $\sim 50-80$ $\rm cts$ $\rm
s^{-1}$ prior to background subtraction (Fig.~2).


The PCA response matrix for this data set was created by adding together
the individual response matrices for PCUs 0, 1, and 2. These individual
matrices were provided by the RXTE Guest Observer Facility (GOF) at Goddard
Space Flight Center and are representative of the most up-to-date PCA
calibration.

Figure~1 shows the background subtracted PCA light curve of our observation
over the whole PCA energy band (2--60\,keV).  Significant variability can
be seen during the $\sim 900$\,ks over which our observations were made.

\begin{figure}
\psfig{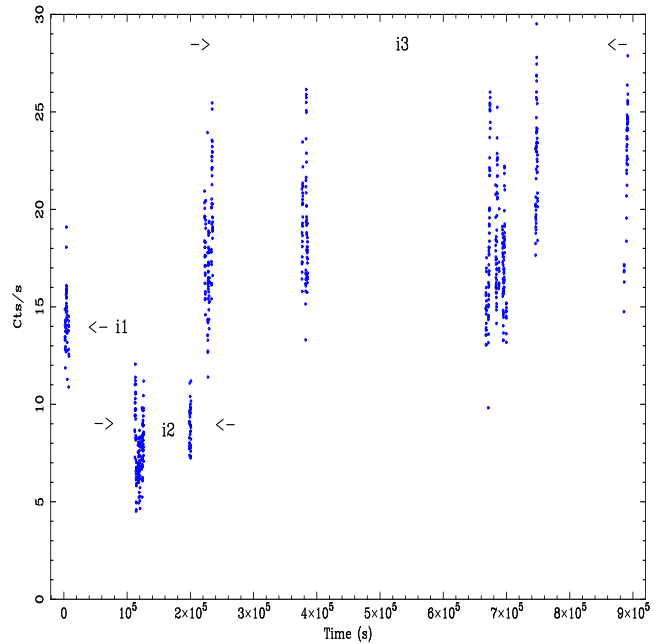}
\caption[h]{Background-subtracted light curve of MCG$-$6-30-15 for observations with 3 PCUs in the 2-60 keV band. The epoch of the start and end time is respectively 1996 September 15, 15:24:17 (UT) and 1996 September 25, 23:08:13 (UT). Time intervals for the three spectral phases of light curve correspond roughly to time intervals 0 - $3 \times 10^3$, $1 \times 10^5$ - $2 \times 10^5$, and $2 \times 10^5$ - $9 \times 10^5$ seconds. }
\end{figure}

\section{Spectral Fitting}

We fit the data in two ways in order to investigate the known features
of reflection and fluorescent iron emission.  A detector Xe feature
previously detected at $\sim 4$ keV leads us to assume that the
possibility of inadequate calibration below this energy may still
exist; uncertainties in calibration certainly still exist below 2 keV.
However, for fairly faint sources such as AGN, background subtraction
is the main source of concern in any {\it RXTE} analysis.  We present
in Fig.~2 light curves showing both the source prior  to background
subtraction as well as the background light curve to give an
indication of the overall background contribution to our observation.
Similarly, a plot of source and background spectra show that
background effects become significant past 20 keV (Fig.~3).  The
standard background subtraction methods described in Section 2.2
should adequately account for the background up until this energy.  We
therefore concentrate on the energy range between 4 and 20 keV for our
fits.  As added checks for the quality of our reduction, background
subtraction and PCA calibrations, we extract spectra from our 400 s
observations of Earth occulted data ($\sc elv
\leq 0$), and a 1.9 ks exposure of archived Crab data from the same gain
epoch (epoch 3) as our observations. We find for the occultation data
that the normalized flux per keV is essentially zero for the
background subtracted source in the energy range of interest.  A power
law fit to the spectrum of the Crab using the most recent response
matrices and background models gives residuals less than 1 per cent
for the energies of interest.  These checks give us extra confidence
in our background subtraction and calibration files.  We add a 1 per
cent systematic error to account for problems with calibration. 

\begin{figure}
\psfig{file=fig2.ps,angle=270,width=8.5truecm,height=8.5truecm}
\end{figure}

\begin{figure}
\psfig{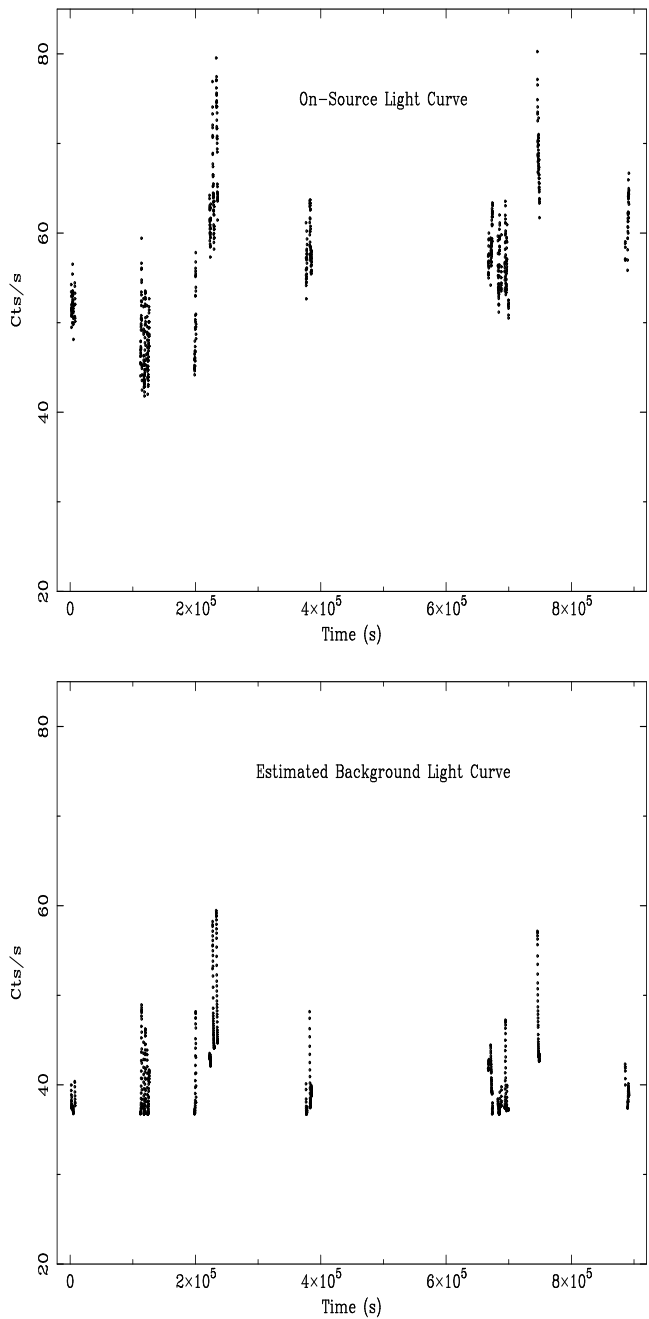}
\caption[h]{Source and background light curve of MCG$-$6-30-15 for observations with 3 PCUs in the 2-60 keV band. The epoch of the start and end time is respectively 1996 September 15, 15:24:17 (UT) and 1996 September 25, 23:08:13 (UT). }
\end{figure}

We also note that restricting our analysis to data above 4\,keV
removes the need to model the photoelectric absorption due to Galactic
ISM material, or the warm absorber that is known to be present in this
object.  Both of these spectral features are only important below
$\sim 2\keV$  (Reynolds et al. 1995; Otani et al. 1996). 

In order to investigate the effects of spectral changes on the photon
index and intensity of the Fe line, we also explore the behavior of
the light curve in its 3 spectral states (Fig.~1) in Section~4. 

\begin{figure}
  \psfig{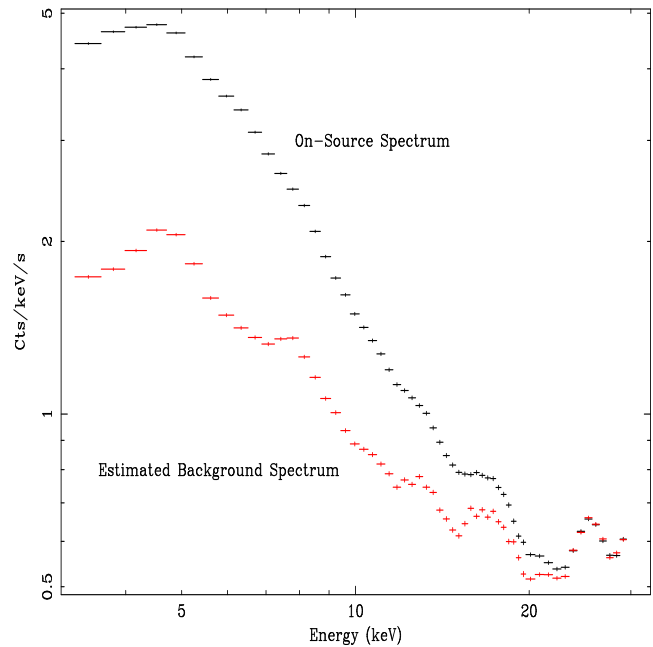}

\caption[h]{Plot of total on-source (i.e. source+background) and background
  spectra in the energy range 3 keV $<$ E $<$ 30 keV show that
  background effects are significant only past 20 keV.} 
\end{figure}
\subsection{Simple Power Law Fits}
A naive fit using a simple power law shows the presence of a
redshifted broad iron line at $\sim$ 6.1 keV and strong reflection
continuum above 10 keV (Fig.~4).  Previous ASCA results have resolved
the presence of a broad iron line in MCG$-$6-30-15 (Fabian et al. 1994a;
Tanaka et al. 1995) and reflection was reported  with GINGA for a
number of Seyfert 1 AGNs (Pounds et al. 1990; Matsuoka et al 1990;
Nandra \& Pounds 1994).  Due to the improved signal-to-noise (as
compared with {\it Ginga}) and higher energy coverage (as compared
with ASCA) afforded by {\it RXTE}, we are able to report here on one
of the first results where the simultaneous strong presence of both
features is evident. 

\begin{figure}
\psfig{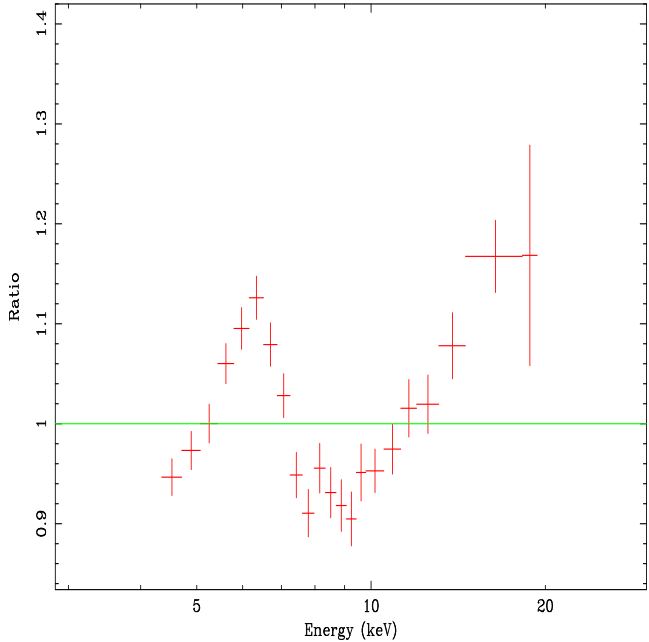}
\caption[h]{Ratio of data to continuum when a simple power law is used to fit
MCG$-$6-30-15 spectra.  There is clear evidence for a redshifted broad
iron line at $\sim$ 6.1 keV and reflection continuum above 10 keV.
This is consistent with  previous GINGA findings for the existence of
a reflection component in many Seyfert 1 AGNs.} 
\end{figure}

\subsection{Fits using power law with reflection and line emission}

The simple power law fit having proven insufficient, we investigate a
multiple component model consisting of a power law, reflection
continuum and gaussian component (to model the iron emission line).
We fit the underlying continuum using the model {\sc pexrav} which is
a power law with an exponential cut off at high energies reflected by
an optically thick slab of  neutral material \cite{mz}.  We fix the
inclination angle of the reflector at $30^\circ$ so as to agree with
the disk inclination one obtains when fitting accretion disk models to
the iron line profile as seen by {\it ASCA} (Tanaka et al. 1995).  For
completeness, we consider the effect of the  $2^\circ -3^\circ$
uncertainty associated with ASCA's determination of the iron line
inclination angle.  This should be a negligible effect since the
reflected component is insensitive to this parameter.  We confirm this
when  we compare contour plots of abundance versus reflective fraction
for inclination angles of $30 \pm 5^{\circ}$ 
and find them to be similar.  The reflective
fraction is defined such that its value equal to unity implies that
the reflecting matter subtends half of the sky ($\Omega = 2 \pi$). 

First, we shall address the formal detectability of the iron line and
the resolution of its width.  Fitting the above continuum model, and
then adding a narrow iron line (with energy as a free parameter) leads
to an improvement in the goodness of fit of $\Delta\chi^2=27$ for two
extra parameters, more than 99 per cent  significant according to the
{\it F-test} for 35 degrees of freedom.  Allowing the line width to be a
free parameter leads to a further improvement of $\Delta\chi^2=7$ for
one additional parameter, with the best fit line width 
$\sigma= 0.55^{+0.29}_{-0.22}$, more than 99 per cent significant according
to the {\it F-test}.
We conclude that the line is both detected and
resolved.  Motivated by the best fit line width, and previous {\it
ASCA} results, we shall fix $\sigma=0.4\keV$ for our future spectral fits.

We find that the best fit is given by a redshifted broad iron line
having an energy of $6.15^{+0.13}_{-0.12}$ keV and equivalent width EW
= $223^{+41}_{-57}$ eV.  With the lighter elemental abundances
(eg. oxygen) set equal to that of iron, we find abundance measurements
to be $0.77^{+4.33}_{-0.25}$ solar abundances with an associated
reflective fraction of $1.43^{+1.23}_{-0.51}$ and spectral index
$\Gamma$ = $2.25^{+0.22}_{-0.09}$ at the 90 per cent confidence
level. At  68 per cent confidence, abundance and reflective fraction
are respectively  $0.77^{+0.39}_{-0.18}$ solar abundances and
$1.44^{+0.57}_{-0.36}$. The overall  $\chi^2 / \nu$ is 0.92 for 35
degrees of freedom.  Figure~5 demonstrates that this model describes
the 4--20\,keV data well (i.e., there are no systematic deviations of
the data from the model).  As expected, there is a strong coupling
between the fit parameters.  In particular, the power-law index, the
elemental abundances and the reflective fraction are strongly coupled.
Figure~6 shows the confidence contours for abundance and reflective
fraction as expected from a corona+disk model. We note that there are
a number of scenarios in which the reflected spectrum can be enhanced
(reflective fraction $>$ 1). These can include effects due to geometry
(ie. where the direct X-ray flares are partially obscured), motion of
the source (e.g. Reynolds \& Fabian 1997), or gravity (light-bending
effects that will beam/focus more of the emission down towards the
disk ; Martocchia \& Matt 1996) in addition to a number of other
possibilities. 

In a fit where we further assume that the primary X-ray source is
above the accretion disk subtending an angle of $2 \pi$ sr (ie
$\frac{\Omega}{2 \pi} = 1$), we find the best fit to be : $\Gamma$ =
$2.15^{+0.1}_{-0.11}$ (this value is slightly steeper than previous
ASCA and Ginga measurements) with the power law flux at 1 keV, $A =
(2.53 \pm 0.36) \times 10^{-2}$ $\rm ph$ $\rm cm^{-2}$ $\rm s^{-1}$,
the redshifted line energy is $6.13 \pm 0.16$ keV with intensity of
the iron line I = $1.47^{+0.18}_{-0.47} \times 10^{-4}$ 
\phpcmsqps and EW = $257^{+32}_{-81}$ eV.
All abundances again set equal to that of iron, we find values for
abundances to be $0.80^{+0.46}_{-0.3}$ and $0.80^{+0.16}_{-0.19}$
solar abundances for 90 and 68 per cent confidence
respectively. $\chi^2 / \nu$ is 1.00 for 36 degrees of freedom. It
should be noted that even though we set the lighter elemental
abundances equal to that of iron, we expect that iron is primary in
determining the fitted abundances since the reflection continuum is
only important at energies past 10 keV where iron absorption
dominates. The significance of iron in affecting abundance values is
confirmed when the lower elemental abundances are decoupled from that
of iron abundances. 

\begin{figure}
\psfig{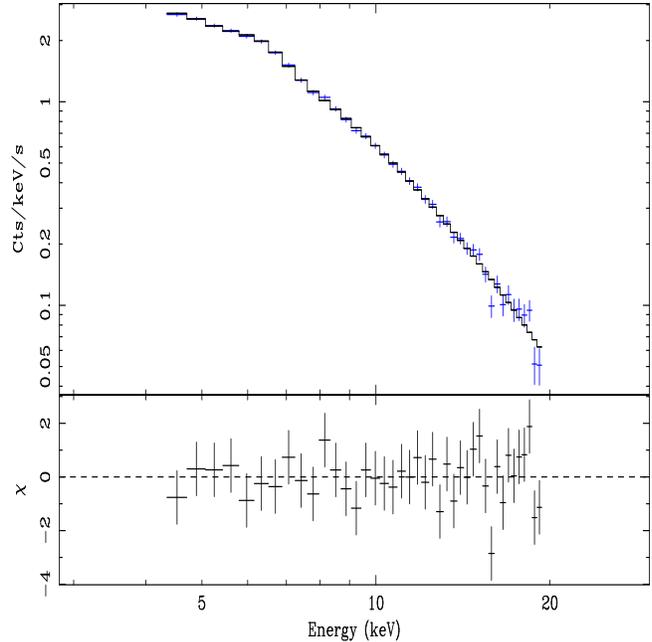}
\caption[h]{A more sophisticated multiple component model fit consisting of 
a gaussian component to represent the iron $K \alpha$ emission ,
and a power law reflection component to model the primary and reflected 
continuum is needed to fit the data well.} 
\end{figure}

The relatively steep intrinsic photon index of MCG$-$6-30-15 is worth
noting.  Previous studies have found a strong anticorrelation between
$\Gamma$ and the full width at  half-maximum (FWHM) of the $H \beta$
emission line for Seyfert 1 type galaxies in  both the soft ROSAT band
(eg. Boller, Brandt \& Fink 1996) and ASCA hard band  (eg. Brandt,
Mathur, \& Elvis 1997).  A recent multiwavelength study of MCG$-$6-30-15
by Reynolds, {\it et al.} 1997 find FWHM of $H \beta$ in the optical
band to be  $2400 \pm 200$ km/s for this object.  Given the steep
intrinsic photon index that  we find for MCG$-$6-30-15, this implies
that MCG$-$6-30-15 could possibily be categorized as a  narrow-line Seyfert 1
galaxy. 

\section{Temporal and Spectral Changes}

Having convinced ourselves that reflection is needed to model our
data, we next investigate the effects of temporal changes in the light
curve on spectral components. To do this, we divide our light curves
into three groups based on the countrate.  Accordingly, we define i1,
i2, and i3 to be respectively the intervals with average count rates
$\sim$ 14, 9, and 20 counts per second.  Figure~1 shows that this also
corresponds roughly to the time intervals 0--3 \,ks, 100--200 \,ks,
and 200--300 \,ks. 

We have searched for spectral changes between the different periods of
data.  From our theoretical pre-conceptions, we also wish to address
whether any changes we see are due to an intrinsic change in the power
law slope, the reflection continuum changing, or a combination of
both. 

Initially, we take a simple approach and fit a simple power law plus
gaussian component in the 4-10 keV energy range, and a simple power
law to the 10-20 keV range.  The 4--10\,keV range should be relatively
unaffected by the reflection component, whereas the 10--20\,keV range
is very much affected by reflection.  These fits are presented in
Table~1.  It can be seen that there is evidence for spectral
variability between these different flux states.  In particular, the
4--10\,keV photon index is significantly steeper in the brightest
period.  In addition, there may be some evidence for a relative
flattening of the 10--20\,keV range in the lowest flux state implying
a possible increase in reflective fraction (this agrees with the more
complicated fit that includes the  reflected spectrum shown in
Table~2), although the errors are too large to make definitive
statements.  The increase of the iron line strength in this low period
would support such a hypothesis. 

\begin{figure}
\psfig{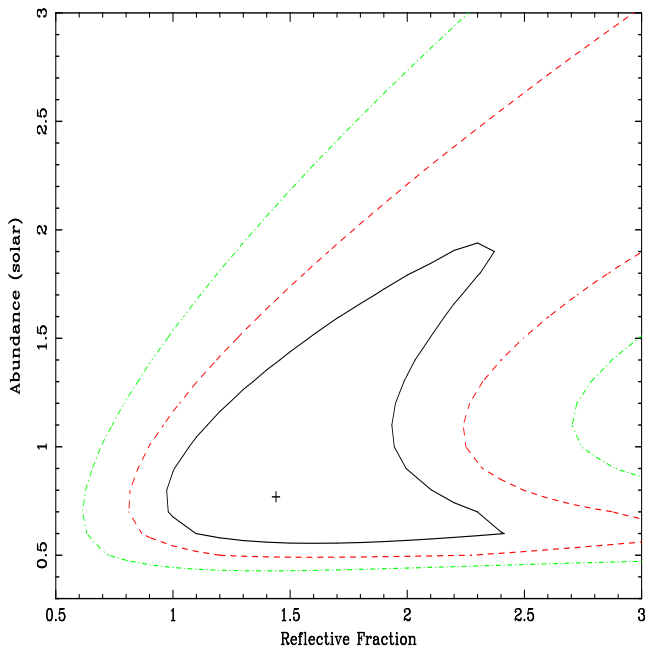}
\caption[h]{Confidence contours (corresponding to 68 per cent, 90 per cent, and 95 per cent confidence levels) for abundance vs. reflective fraction relationship for fits in the energy range 4 keV $<$ E $<$ 20 keV.}
\end{figure}

\begin{table}
\begin{center}
\begin{tabular}{|c|c|c|c|c|}
\multicolumn{5}{c}{\sc Change in $\Gamma$ for different states of MCG$-$6-30-15} \\
\hline
{\em \rm Interval} & {\em $\rm \Gamma_{4-10}^a$} & {\em $\rm
\Gamma_{10-20}^b$} & {\em $\rm W^c$ (eV) } & {\em $\rm Flux^d$ }  \\
\hline
\hline
i1 & $1.89^{+0.1}_{-0.11}$ & $1.53^{+0.32}_{-0.36}$ &
$327^{+112}_{-80}$ & 4.06 \\  i2 & $1.86^{+0.09}_{-0.1}$ &
$1.29^{+0.22}_{-0.35}$ & $446^{+85}_{-82}$ &  2.51 \\  i3 &
$2.10^{+0.05}_{-0.04}$ & $1.69^{+0.13}_{-0.12}$ & $253^{+35}_{-48}$ &
6.03 \\ 
\hline

\end{tabular}
\caption{Results are quoted from simple power law fits. 
$^a$ Power-law photon index in for 4 keV $<$ E $<$ 10 keV.  $^b$
Power-law photon index for 10 keV $<$ E $<$ 20 keV.  $^c$ Equivalent
width of the iron emission line.  $^d$ 2-10 keV flux in units of
$10^{-11}$ \ergpcmsqps}

\end{center}
\end{table}

\begin{table*}
\begin{center}
\begin{tabular}{|c|c|c|c|c|c|c|c|c|}
\multicolumn{9}{c}{\sc Spectral Fits using power-law with reflection model for different states of MCG$-$6-30-15}\\
\hline
{\em \rm Data} & {\em $\rm \Gamma_{4-20}^a$} & {\em $\rm A^b$ } & {\em
$\rm refl^c$ } & {\em $\rm LineE^d$ } & {\em $\rm I_{K \alpha}^e$} &
{\em $\rm W^f$ (eV)} & {\em $\rm Flux^g$ } & {\em $\rm \chi^2$}  \\
\hline
\hline
i1 & $2.01^{+0.25}_{-0.19}$ & $1.61^{+0.58}_{-0.39}$ &
$1.55^{+2.29}_{-2.03}$ & $5.99^{+0.23}_{-0.21}$  &
$1.44^{+0.45}_{-0.64}$ & $283^{+88}_{-126}$ & 6.83 & 21 \\  i2 &
$2.04^{+0.22}_{-0.18}$ & $1.01^{+0.34}_{-0.22}$ &
$1.95^{+2.36}_{-1.08}$ & $6.17^{+0.15}_{-0.14}$ &
$1.11^{+0.23}_{-0.34}$ & $374^{+77}_{-115}$ & 4.26 & 23 \\  i3 &
$2.22^{+0.08}_{-0.08}$ & $3.45^{+0.44}_{-0.37}$ &
$1.28^{+0.56}_{-0.42}$ & $6.15^{+0.08}_{-0.13}$ &
$1.26^{+0.34}_{-0.27}$ & $182^{+49}_{-39}$ &  9.12 & 32 \\ 
\hline

\end{tabular}
\caption{$^a$ Power-law photon index. $^b$ Power-law flux at 1 keV, in units of $10^{-3}$ $\rm ph $ $\rm cm^2$ $\rm s^{-1}$ $\rm keV^{-1}$. $^c$ Reflective fraction = $\Omega / 2\pi$. $^d$ Energy of the iron $K \alpha$ emission line. $^e$ Intensity of iron emission line in units of $10^{-4}$ \phpcmsqps . $^f$ Equivalent width of the emission line. $^g$ 2-20 keV flux in units of $10^{-11}$ \ergpcmsqps. $^h$ $\chi^2$ for 36 degrees of freedom}

\end{center}
\end{table*}

We next use the {\sc pexrav} model to investigate more rigorously the
relationship between the spectral components, $\Gamma$ and the reflective
fraction.  This is an important issue to address since it has direct
bearing on the physics of the central engine.  For example, if the
spectral variability can be shown to originate purely from a change in
the photon index of the primary source (with constant reflective
fraction), changes in the conditions of the X-ray emitting disk-corona
would then be strongly implicated.  On the other hand, the different
temporal states may be due to the amount of reprocessing in the
source; for example, the gravitational bending/focusing as the X-ray
source gets closer to the black hole will enhance the amount of
reflection (e.g., Martocchia \& Matt 1996).  Fig.~7 shows the
confidence contours on the reflective-fraction/photon-index plane for
these three periods of data. To improve our constraints, the
abundances of the reflector have been fixed at the best-fit value
found from the total dataset ($Z=0.77{\rm Z}_\odot$; Section 3.2).  We
are justified in fixing these abundance since, on physical grounds, we
do not expect them to vary between our three intervals.  Given that we
have fixed the elemental abundances, we also fixed the ratio between
the iron line equivalent width and the reflective fraction such that a
reflective fraction of unity corresponds to an iron line equivalent
width of 150\,eV (appropriate for the above choice of abundances; see
Reynolds, Fabian \& Inoue 1995). 

As suspected on the basis of the simple power-law fits in Table~1,
Fig.~7 (using a more complicated fit that includes the reflected
spectrum) shows that the steepening in primary photon index during the
brightest period of data is statistically significant.  There is also
a suggestion of an increase in reflective fraction (with constant
$\Gamma$) when going from the intermediate flux state to the low flux
state.  Thus it appears that both changes in the properties of the
primary X-ray source and changes in the amount of reflection are
relevant for understanding spectral variability. As a further check on
the robustness of these results, we have recreated contours similar to
those in Fig.~7 for several choices of abundances within the 68 per
cent confidence statistical error range, and find that none of the
results changes materially based on these choices of abundance values.
The  discrepancy in $\Gamma$ between i3 and the other two states does
diminish if the abundance is allowed to be a free parameter.  It
should be noted that the model as currently implemented does not
include relativistic blurring of  the reflection component. A major
caveat to RXTE spectral variability results lies in the background
models.

\begin{figure}
\psfig{file=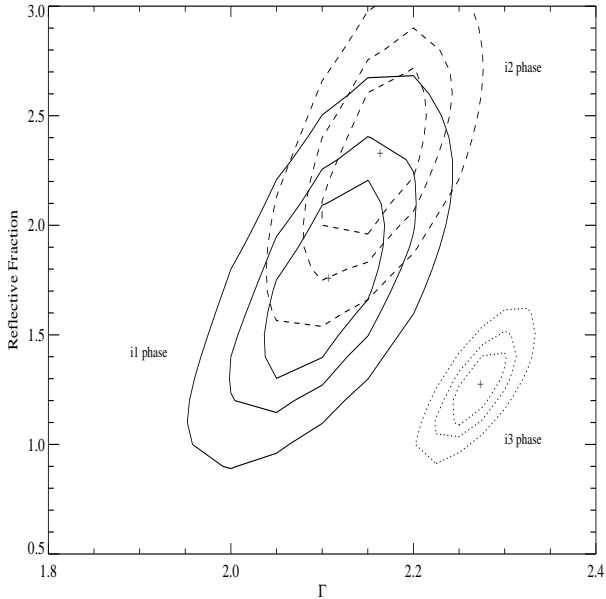,angle=90,width=8.5truecm,height=8.5truecm}
\caption{Confidence contours (corresponding to 68 per cent, 90 per cent, and 95 
per cent confidence levels) showing the relationship between $\Gamma$
and reflec tive fraction for i1, i2, and i3 phases of MCG$-$6-30-15
light curve during the period from 1996 September 15 to 1996
September 25.} 
\end{figure}

\section{Discussion}

The purpose of this paper was to show what unanswered questions can be
addressed with the large area and wide-band coverage of {\it RXTE}
even with the current uncertainties in spectral calibration. The
presence of a broad iron line is clearly evident as shown with a
simple power law fit, and is one of the first detections where both
features are seen simultaneously.  We add a reflection component to
our power law and gaussian fit to find that reflection is necessary to
describe our data.  We note also that the steep intrinsic photon index
coupled with a narrow $H \beta$ FWHM implies that MCG$-$6-30-15 can be a
possible narrow-line Seyfert 1 galaxy candidate. 

While spectral results may change in detail over the course of the
next year with further improvements in calibration, we can already
begin to place upper bound limits on the relationship between
abundance values and reflective fraction. 

In Section 4, we study the effects of temporal variability on spectral
components and find evidence to support the notion that variability
may be due to changes in the amount of reflection seen (e.g. due to
gravitational or Doppler beaming of the primary emission towards the
disk).  It is however not clear whether this effect may also be
coupled with  contributions from changes in the properties of the
source itself (e.g. the temperature and optical depth of the coronal
plasma).  We expect to be able to resolve these issues better with
longer looks and simultaneous ASCA observations.  For the time being,
the present results are important observational first  steps in
understanding some of the physics of AGN reprocessing mechanisms, and
push the limits of our knowledge. 

\section{ACKNOWLEDGEMENTS}
We thank all the members of the RXTE GOF for answering our inquiries
in such a timely manner, with special thanks to Keith Jahoda for
explanations of calibration issues.  JCL thanks the Isaac Newton
Trust, the Overseas Research Studentship programme (ORS) and the
Cambridge Commonwealth Trust for support.  ACF thanks the Royal
Society for support. CSR thanks the National  Science Foundation for
support under grant AST9529175, and NASA for support under the Long
Term Space Astrophysics grant NASA-NAG-6337. KI and WNB thank PPARC
and NASA RXTE grant NAG5-6852 for support respectively.

\end{document}